\documentclass[10pt,twoside]{article}
\usepackage[tbtags]{amsmath}
\usepackage{accents}
\usepackage{rotating}
\usepackage{amsfonts}
\usepackage{amssymb}
\usepackage{latexsym}

\DeclareMathSymbol{\vecarrow}{\mathord}{letters}{"7E}

\newlength{\lvech}
\newlength{\lvecw}

\newcommand{\lc}[1]{\underset{#1}{\lrcorner}}
\newcommand{\lvec}[1]{\ensuremath{
\text{\settoheight{\lvech}{$\vecarrow$}\addtolength{\lvech}{-.047ex}\settowidth{\lvecw}{$\vecarrow$}$\accentset{\hspace{.47\lvecw}\begin{rotate}{180}\makebox[0pt]{\raisebox{-\lvech}[0pt][0pt]{$\vecarrow$}}\end{rotate}}{#1}$}}}

\renewcommand{\theequation}{arabic {section}.\arabic{equation}}

\newcommand{\starSU}{*_{\scriptscriptstyle SU}}

\newcommand{\starMP}{*_{\scriptscriptstyle MP}}
\newcommand{\starM}{*_{\scriptscriptstyle M}}
\newcommand{\starP}{*_{\scriptscriptstyle P}}
\newcommand{\tstar}{\tilde{\mbox{$\ast$}}}
\newcommand{\tstarM}{\,\tstar_{\scriptscriptstyle M}\,}

\newcommand{\fett}[1]{\mbox{\boldmath$#1$}} \newcommand{\psl}{p\!\!\!/} \newcommand{\usl}{u\!\!\!/} \newcommand{\Su}{S_{\mbox{\scriptsize \boldmath$u$}}}

\newcommand{\beq}{\begin{equation}}
\newcommand{\eeq}{\end{equation}}

\renewcommand{\theequation}{\arabic {section}.\arabic{equation}}

\begin{document}
\makeatletter
\title{Cliffordization, Spin and Fermionic Star Products}
\author{Allen C. Hirshfeld\footnote{hirsh@physik.uni-dortmund.de},\,\,
Peter Henselder\footnote{henselde@dilbert.physik.uni-dortmund.de}\, and
Thomas Spernat\footnote {tspernat@zylon.physik.uni-dortmund.de} \\
Fachbereich Physik, Universit\"at Dortmund\\
44221 Dortmund}

\maketitle
\begin{abstract}

Deformation quantization is a powerful tool for quantizing
theories with bosonic and fermionic degrees of freedom. The star
products involved generate the mathematical structures which have
recently been used in attempts to analyze the algebraic properties
of quantum field theory. In the context of quantum mechanics they
provide a canonical quantization procedure for systems with either
bosonic of fermionic degrees of freedom. We illustrate this
procedure for a number a physical examples, including bosonic,
fermionic and supersymmetric oscillators. We show how
non-relativistic and relativistic particles with spin can be
naturally described in this framework.
\end{abstract}

\section{Introduction}

\qquad The {\it deformation quantization} formulation of quantum
physics was initiated by Bayen et.\ al.\ in Ref.~\cite{Bayen}; for
recent reviews see e.g.\ \cite{Stern,HH}. Physical applications of
this formalism have mainly been restricted to systems involving
bosonic degrees of freedom, e.g.\ \cite{pert}. For some time it
remained unclear how spin and relativistic quantum mechanics could
be described in this formalism. J.~Varilly et.\ al.\ adressed this
problem using Moyal products \cite{Varilly}, in \cite{Varilly2}
they combined their methods with group theoretic arguments in
order to treat particles with spin and the covariance of the Dirac
equation. In contrast to such approaches we advocated in
\cite{ferm} using for systems involving fermionic degrees of
freedom a method based on the work of Berezin and Marinov
\cite{Ber} in which one starts from a pseudoclassical system
described in terms of Grassmann variables, and achieves
quantization by use of a fermionic star product, which appeared in
another context in \cite{Zachos}. We showed how the familiar {\it
Clifford algebra} structures characteristic of particles with spin
arise in this framework.

A more general program for analyzing relativistic quantum field
theories using Clifford algebra structures has been pursued in
recent years by Fauser \cite{Fauser1}-\cite{Fauser6}. In this
approach these structures are derived from an underlying {\it
Grassmann algebra} by the procedure of {\it Chevalley
Cliffordization} \cite{Chev}. In \cite{Fauser6} Fauser discussed
the Dirac equation in this context.

Another program for understanding the algebraic structures which
arise in perturbative quantum field theory has been advocated by
Brouder and Oekl \cite{Oeckl,BO}; the fundamental algebraic
structure they use is the {\it circle product}, introduced by
Brouder in \cite{Brouder}. In \cite{Deform4} we have discussed the
relation of this product to the {\it star products} used in
deformation quantization.

In the present paper we attempt to clarify the relations between
these different approaches to quantum physics. We believe the
deformation quantization approach to be the most fundamental,
relying as it does on Gerstenhaber's seminal analysis of
associative algebras \cite{Gerst}. As soon as the appropriate star
product has been determined the analysis of a quantum mechanical
system proceeds in a canonical fashion: the states are
characterized by the relevant Wigner functions and the eigenvalues
of the Hamiltonian follow from the $*$-genvalue equation or from the
star exponential.

The paper is structured as follows. In Sec.\ 2 we briefly review
the Chevalley Cliffordization procedure for constructing a {\em
Clifford algebra} from an underlying Grassmann algebra. In Sec.\ 3
we elucidate the relationship between Brouder's {\em circle
product} and the Clifford algebra structure of fermionic systems.
Sec.~4 clarifies the relation between Fauser's concept of {\em
Wick isomorphism} and the c-equivalence of deformation
quantization. In Sec.\ 5 we lay out the general scheme for
quantizing a given physical system, both for bosonic and for
fermionic degrees of freedom. We illustrate the method for the
bosonic and fermionic oscillators. We show that the fermionic
angular momentum corresponds to the usual spin concept of
non-relativistic quantum mechanics, and verify the basic
properties of the spin vector. Sec.\ 6 treats a real physical
system: a charged particle in a constant magnetic field. For a
spinless particle the well-known Landau energy levels are
recovered, as well as the eigenfunctions of the orbital angular
momentum. For a particle with spin one-half the system corresponds
to a {\em supersymmetric} oscillator, with its characteristic
degenerate energy levels. Sec.\ 7 shows how the determination of
exact and broken supersymmetry in terms of the {\em Fredholm} or
{\em Witten index} follows in this formalism. Sec.\ 8 discusses
different representations for the {\em Dirac operators} in this
context. We determine the relevant Wigner functions and the star
exponential. We also exhibit the star product analogues of the
Dirac spin projectors. In Sec.\ 9 we follow the Foldy-Wouthuysen
procedure in order to study the non-relativistic approximation to
the Dirac equation and show how the conventional operator
expressions may be recovered by use of the {\em Weyl transform}.
Sec.\ 10 contains our conclusions, and an outlook for further
research.

\section{Chevalley Cliffordization}
\setcounter{equation}{0}\label{ChevCliff}

\qquad In this section we briefly review the construction of a
Clifford algebra from a Grassmann algebra. This subject was
developed by Cartan, Weyl and Chevalley \cite{Chev}. We follow the
notation of Fauser \cite{Fauser1}.

The starting point is a Grassmann algebra $Gr$. This is the
$\mathbb{Z}$-graded algebra generated by a set of Grassmann
variables $\{\theta_1,\ldots \theta_n\}$, which satisfy the
relations
\begin{equation}
\theta_i \theta_j=-\theta_j \theta_i\ \ \ \ \forall i,j=1\ldots n.
\end{equation}
We also take as given a bilinear form
\begin{equation}
B(\theta_i,\theta_j)=g(\theta_i,\theta_j)+A(\theta_i,\theta_j),
\end{equation}
where $g$ and $A$ are the symmetric and antisymmetric parts, respectively.

We define an antiderivation on $Gr$ as a map which acts on generators and
monomials according to the following rules:
\begin{subequations}
\begin{eqnarray}
\theta_i\lc{B}\theta_j & = & B(\theta_i,\theta_j),\label{i}\\
\theta_i\lc{B}\left( u v\right) & = &
(\theta_i\lc{B}u) v+(-1)^{\pi(u)}u(\theta_i\lc{B}v) \label{ii}\\
\left( u v\right)\lc{B}w & = & u\lc{B}(v\lc{B}w). \label{iii}
\end{eqnarray} \label{Regeln}\end{subequations}Here $u$ and $v$ are homogeneous monomials, and $\pi(u)$ is the
grade of $u$. The mapping is then linearly extended to arbitrary
elements of $Gr$. From Eq.~(\ref{ii}) with $u=v=1$ it is clear
that $ \theta_i\lc{B}1=0$. From Eq.~(\ref{iii}) with $v=1$ it
follows that $ 1\lc{B}u=u. $ For homogenous $u$ and $v$ we have
\begin{equation}
\pi(u\lc{B}v)= \pi(v)-\pi(u) \label{Graduv}.
\end{equation}

We now define the linear mapping
\begin{eqnarray}
\gamma_{\theta_i}^B: \bigg\{ \raisebox{-0.15pc}{$\displaystyle
\begin{array}{rl}
Gr~\rightarrow & \hspace{-0.3pc} Gr \\
 u~\mapsto     & \hspace{-0.3pc} \theta _i u + \theta _i \lc{B} u \ .
\end{array}$}\end{eqnarray}We easily calculate
\begin{equation}
\gamma_{\theta_i}^B\gamma_{\theta_j}^Bu=
B(\theta_i,\theta_j)u+(\theta_i \theta_j)\lc{B}u+\theta_i\theta_ju
  +\big[\theta_i(\theta_j\lc{B}u)-\theta_j(\theta_i\lc{B}u)\big].
\label{gammagamma}
\end{equation}
From this we see that the $\gamma_i$ are the generators of a
Clifford algebra $Cl(B)$, since
\begin{equation}
\{ \gamma_{\theta_i}^B \gamma_{\theta_j}^B \}:=
\gamma_{\theta_i}^B \gamma_{\theta_i}^B +\gamma_{\theta_i}^B
\gamma_{\theta_i}^B = 2 g(\theta_i,\theta_j). \label{anticom}
\end{equation}

\section{Circle Products for Grassmann Variables}
\setcounter{equation}{0}\label{circle}

\qquad By now there are a number of associative products in the
literature which are used to discuss the algebraic structure of
quantum mechanics and quantum field theory. In Ref.\
\cite{Deform4} we discussed the relation between Brouder's circle
product \cite{Brouder}, which is a special case of Drinfeld's
twisted product \cite{Drinfield}, and the star product of
deformation quantization. In the present section we discuss the
relation of the circle product to the product encountered in the
Chevalley Cliffordization procedure.

The fermionic version of the circle product \cite{ferm} is
\begin{equation}
u\circ_B v= u\, \mathrm{exp}\left(\sum_{i,j}B(\theta_i,\theta_j)
\lvec{\partial}_{\theta_i}\vec{\partial}_{\theta_j}\right)\, v.
\label{circdef}
\end{equation}
In the above formula the arrows indicate on which function the
differential operators are acting. The differential operators
which act to the right are left derivatives, those which act to
the left are right derivatives with respect to the Grassmann
variables. The following discussion is only valid for monomials,
but the generalization to arbitrary elements of $Gr$ is
straightforward. Since the $n$-th term in the expansion of
$u\circ_B v$ is of grade $\pi(u)+\pi(v)-2n$ one can compare the
$\pi(u)$-th term with $u\lc{B}v$, which is of the same grade:
$\pi(v)-\pi(u)$. In fact, both turn out to be identical, i.\,e.
\begin{eqnarray}
u\lc{B}v&=&\frac{1}{\pi(u)!}\,u\left(\sum_{i,j}B(\theta_i,\theta_j)
\lvec{\partial}_{\theta_i}
            \vec{\partial}_{\theta_j}\right)^{\pi(u)}\,v\nonumber \\
&=&u\left(\sum_ {\sum k_{ij}=\pi(u)}
     \prod_{i,j=1}^n \left( B(\theta_i,\theta_j)\lvec{\partial}_{\theta_i}
       \vec{\partial}_{\theta_j}\right)^{k_{ij}}
      \right)v,
\label{hakdar}
\end{eqnarray}
where the $k_{ij}$ are either 1 or 0. To prove this equality we
have to show that the three axioms of (\ref{Regeln}) are
fulfilled. The first axiom is trivial, the second one follows from
the Leibniz rule
\begin{eqnarray}
\theta_i\lc{B}(uv)&=&\sum_jB(\theta_i,\theta_j)\vec{\partial}_{\theta_j}(uv)
\nonumber\\
&=&\sum_jB(\theta_i,\theta_j)\left[(\partial_{\theta_j}u)v+(-1)^{\pi(u)}u(
\partial_{\theta_j}v)
\right]\nonumber\\
&=&(\theta_i\lc{B}u)v+(-1)^{\pi(u)}u(\theta_i\lc{B}v)
\label{hakleib}
\end{eqnarray}
and a proof of (\ref{iii}) can be found in the appendix. Therefore
$u\lc{B}v$ is equal to the term of the expansion of $u\circ_B v$
in which all basis elements of $\theta_i$ in $u$ are cancelled by
corresponding derivatives $\lvec \partial_{\theta_i}$. Such a term
will only exist if $\pi(u)\le\pi(v)$ and if the necessary
derivatives appear, i.e.\ the corresponding $B(\theta_i,\theta_j)$
have to be non-zero.

One can now formulate the Clifford map with the help of a circle
product as
\begin{equation}
\gamma_{\theta_i}^Bu=\left(\theta_i+\theta_i\lc{B}\right)u=\theta_i\circ_B
u. \label{gammacirc}
\end{equation}
We generalize our previous notation and write for general
homogeneous $u$ and $v$
\begin{equation} \gamma_v u=v\circ_B u,
\end{equation}
which implies
\begin{equation}
\gamma_u\gamma_v=\gamma_{u\circ v}.
\end{equation}
With this
notation Eq.\ (\ref{gammagamma}) reads
\begin{eqnarray}
\theta_i\circ_B \theta_j\circ_B u
&=&\theta_i\theta_ju+\sum_{k,l}B(\theta_j,\theta_k)B(\theta_i,\theta_l)
\vec{\partial}_{\theta_l}
\vec{\partial}_{\theta_k}u+B(\theta_i,\theta_j)u\nonumber\\
&&+\theta_i\sum_kB(\theta_j,\theta_k)\vec{\partial}_{\theta_k}u
-\theta_j\sum_lB(\theta_i,\theta_l)\vec{\partial}_{\theta_l}u
\end{eqnarray}
and the anticommutator (\ref{anticom}) can be written as
\begin{equation}
\{\theta_i,\theta_j\}_{\circ_B}\circ_B u=
\{\theta_i,\theta_j\}_{\circ_B}=\theta_i\circ_B\theta_j+\theta_j\circ_B\theta_i
=2g(\theta_i,\theta_j)u\text{.}\label{antisym}
\end{equation}

\section{The Wick Isomorphism and c-Equivalence}
\setcounter{equation}{0}\label{c-eq}

\qquad In quantum mechanics and quantum field theory the Clifford
algebras $C\ell(g)$ and $C\ell(B)$ are related. Fauser
 \cite{Fauser1} uses the concept of {\em Wick isomorphism} to
 express this relationship in terms of  the grade-2 form
$F=F^{ij}\theta_i\theta_j $, which is related to the antisymmetric
part of $B$:
\begin{equation}
\sum_{r,s}F^{rs}g(\theta_i,\theta_s)g(\theta_j,\theta_r)=\frac{1}{2}A(\theta_i,\theta_j).
\label{F}
\end{equation}
The Wick isomorphism maps a monomial $u$ into $e^{-F}\,u\,e^F.$ In
this section we shall discuss this relationship in terms of circle
products.

We start by calculating
\begin{equation}
\theta_i\lc{B}F=\sum_{j}B(\theta_i,\theta_j)\vec{\partial}_{\theta_j}F^{kl}
\theta_k\theta_l=\sum_j 2B(\theta_i,\theta_j)F^{jk}\theta_k.
\end{equation}
This leads to
\begin{equation}
\theta_i\lc{B}F^n= n(\theta_i\lc{B}F)F^{n-1},
\end{equation}
which implies
\begin{equation} \theta_i\lc{B}e^F=
(\theta_i\lc{B}F)e^F,
\end{equation}
so that we find
\begin{equation}
e^{-F}\left[\theta_i\lc{B}(e^Fu)\right]=
\theta_i\lc{B}u+(\theta_i\lc{B}F)u.
\end{equation}
With this we calculate
\begin{equation}
e^{-F}\gamma_{\theta_i}^{g}e^Fu=\theta_iu+\theta_i\lc{g}u+(\theta_i\lc{g}F)u.
\label{eingamma}
\end{equation}
Similarly we find
\begin{eqnarray}
e^{-F}\gamma_{\theta_i}^g\gamma_{\theta_j}^ge^Fu
&=&\theta_i\theta_ju+g(\theta_i,\theta_j)u
+\theta_i(\theta_j\lc{g}F)u-\theta_j(\theta_i\lc{g}F)u\nonumber\\
&&+\theta_i(\theta_j\lc{g}u)-\theta_j(\theta_i\lc{g}u)
+(\theta_i\lc{g}F)(\theta_j\lc{g}u)-(\theta_j\lc{g}F)(\theta_i\lc{g}v)
+\theta_i\lc{g}(\theta_j\lc{g}u)-(\theta_j\lc{g}F)(\theta_i\lc{g}F)u\nonumber\\
&&+(\theta_i\lc{g}(\theta_j\lc{g}F))u.
\label{witra}
\end{eqnarray}
In this expression there are two terms that multiply $u$ by a
scalar, namely $g(\theta_i,\theta_j)u$ and
$(\theta_i\lc{g}(\theta_j\lc{g}F))u$. This last term is
\begin{eqnarray}
\theta_i\lc{g}(\theta_j\lc{g}F)&=&\theta_i\lc{g}\left(\sum_k g(\theta_j,
\theta_k)\vec{\partial}_{\theta_k}F^{rs}\theta_r\theta_s\right)\nonumber\\
&=&\theta_i\lc{g}\left(2\sum_r g(\theta_j,\theta_r)F^{rs}\theta_s\right)
\nonumber\\
&=&2\sum_{r,s}F^{rs}g(\theta_i,\theta_s)g(\theta_j,\theta_r)=A(\theta_i,
\theta_j) \label{Aexpli}.
\end{eqnarray}
Hence the Wick isomorphism has induced an antisymmetric term
$A(\theta_i, \theta_j)$ that combines with the symmetric term
$g(\theta_i,\theta_j)$ to $B(\theta_i,\theta_j)$. By symmetrizing
Eq.\ (\ref{witra}) in $i$ and $j$ one sees that the anticommutator
is invariant with respect to the Wick isomorphism:
\begin{equation}
e^{-F}\{\gamma_{\theta_i}^{g},\gamma_{\theta_j}^{g}\}e^F
=2g(\theta_i,\theta_j)\text{.}
\end{equation}

The concept of Wick isomorphism is similar to the concept of
$c$-{\em equivalence} in the context of star products. Two star
products are $c$-equivalent if they are related by a
$T$-transformation:
\begin{equation}
u\circ'v=T^{-1}(Tu\circ Tv)
\end{equation}
with $T=\mathrm{exp}\big(T^{ij}\vec{\partial}_{\theta_i}
\vec{\partial}_{\theta_j}\big)$. Transforming the Clifford maps
into the circle product notation as discussed in the last section
we see that the Wick isomorphism does not transform $\circ_g$ into
$\circ_B$, as a $T$-transformation would. This can be seen from
the simple fact that for $u=1$ Eq.\ (\ref{witra}) leads to
\begin{eqnarray}
e^{-F}\left(\gamma_{\theta_i}^g\gamma_{\theta_j}^ge^F\right)
&=&e^{-F}\left(\theta_i\circ\theta_j\circ e^F\right)\nonumber\\
&=&\theta_i\theta_j+g(\theta_i,\theta_j)+(\theta_i\lc{g}F)\theta_j
+\theta_i(\theta_j\lc{g}F)+(\theta_i\lc{g}F)(\theta_j\lc{g}F)
+\theta_i\lc{g}(\theta_j\lc{g}F),
\end{eqnarray}
where a number of terms of order two appear, while in
$\theta_i\circ_B\theta_j$ the only term of order two is
$\theta_1\theta_2$. So the Wick isomorphism does not lead to a
$T$-transformation of the corresponding circle product. But as it
does induce an antisymmetric scalar part, and this scalar part is
just the scalar part of the $T$-transformed circle product. The
following result holds true:
\begin{equation}
\varepsilon\left[\theta_{i_1}\circ_B\cdots\circ_B\theta_{i_n}\right]
=
\varepsilon\left[e^{-F}\left(\theta_{i_1}\circ_g\cdots\circ_g\theta_{i_n}
\circ_g e^F \right)\right], \label{theo}
\end{equation}
where $\varepsilon$ projects onto the scalar part of the
expression. In terms of Clifford algebras this means that although
$e^{-F}C\ell(g,V) e^{+F}$ is not equal to $C\ell(B,V)$, the
equation
\begin{equation}
\varepsilon \left[e^{-F}C\ell(g,V) e^{+F}\right]=\varepsilon
\left[C\ell(B,V)\right]
\end{equation} is valid.

For $n$ odd the relation (\ref{theo}) is empty; both sides of the
equation vanish. For even $n$ the left hand side yields
\begin{eqnarray}
\varepsilon\left[\theta_{i_1}\circ_B\cdots\circ_B \theta_{i_{2m}}\right]
&=&\sum_{\sigma\in S_{2m}}(-1)^{\sigma}B(\theta_{\sigma(i_1)},
\theta_{\sigma(i_2)})
\cdots B(\theta_{\sigma(i_{2m-1})},\theta_{\sigma(i_{2m})})\nonumber\\
&=&\sum_{\sigma\in S_{2m}}(-1)^{\sigma}
(g(\theta_{\sigma(i_1)},\theta_{\sigma(i_2)})+A(\theta_{\sigma(i_1)},
\theta_{\sigma(i_2)}))
\nonumber\\
&&\qquad \qquad\cdots(g(\theta_{\sigma(i_{2m-1})},\theta_{\sigma(i_{2m})})
+A(\theta_{\sigma(i_{2m-1})},\theta_{\sigma(i_{2m})}))\nonumber\\
&=&\sum_{\sigma\in S_{2m}}(-1)^{\sigma}\sum_{X=g,A}
X(\theta_{\sigma(i_1)},\theta_{\sigma(i_2)}) \cdots
X(\theta_{\sigma(i_{2m-1})},\theta_{\sigma(i_{2m})}), \label{Wick}
\end{eqnarray}
where we have used the Wick theorem as in Ref.\ \cite{pert}. The
right hand side of Eq.\ (\ref{theo}) involves the term
\begin{eqnarray}
\theta_{i_1}\circ_g\cdots\circ_g \theta_{i_{2m}}&=&\theta_{i_1}\cdots
\theta_{i_{2m}}
+\sum_{\sigma\in S_{2m}}(-1)^{\sigma}\left[
g(\theta_{\sigma(i_1)},\theta_{\sigma(i_2)})\theta_{\sigma(i_3)}\cdots
\theta_{\sigma(i_{2m})}\right.\nonumber\\
&&\qquad\qquad\qquad\qquad
+g(\theta_{\sigma(i_1)},\theta_{\sigma(i_2)})g(\theta_{\sigma(i_3)},
\theta_{\sigma(i_4)})
\theta_{\sigma(i_5)}\cdots \theta_{\sigma(i_{2m})}\nonumber\\
&&\qquad\qquad\qquad\qquad
\left. +\cdots+g(\theta_{\sigma(i_1)},\theta_{\sigma(i_2)})\cdots
g(\theta_{\sigma(i_{2m-1})},\theta_{\sigma(i_{2m})})\right],
\label{Bew1}
\end{eqnarray}
where we have again used the Wick theorem. For the first term in
this expression we find
\begin{equation}
\varepsilon\left[(\theta_{i_1}\cdots \theta_{i_{2m}})\circ_g
e^F\right] = \sum_{\sigma\in S_{2m}}(-1)^{\sigma}
A(\theta_{\sigma(i_1)}, \theta_{\sigma(i_2)})\cdots
A(\theta_{\sigma(i_{2m-1})},\theta_{\sigma(i_{2m})}), \label{Bew2}
\end{equation}
where we have used the definition of $F$ in terms of $A$, Eq.\
(\ref{F}). Continuing in this way we find
\begin{equation}
\varepsilon\left[e^{-F}(\theta_{i_1}\circ_g\cdots\circ_g
\theta_{i_{2m}} \circ_g e^F) \right] = \sum_{\sigma\in
S_{2m}}(-1)^{\sigma}\sum_{X=g,A}
X(\theta_{\sigma(i_1)},\theta_{\sigma(i_2)}) \cdots
X(\theta_{\sigma(i_{2m-1})},\theta_{\sigma(i_{2m})}),
\end{equation}
which finishes the proof.

The result we have established in this section implies that
although the result of a Wick isomorphism and a $T$-transformation
on a circle product are not identical, their scalar parts are
identical. The scalar parts correspond in quantum field theory to
the vacuum expectation values \cite{Deform4}. This is sufficient
to establish the equivalence of the two procedures in perturbative
quantum field theory, where the relevant quantities are the vacuum
expectation values of products of field operators. For these
vacuum expectation values we see that the choice of the
antisymmetric part of the bilinear form $B$ is of no physical
consequence.

\section{The Quantization of Bosonic and Fermionic Systems}
\setcounter{equation}{0}\label{BFosc} \qquad In this section we
want to show how different specializations of the circle product
(\ref{circdef}) can be used for physical applications. We first
consider a dynamical system involving bosonic degrees of freedom.
The relevant product is then the Moyal product:
\begin{equation}
f\starM g = f\exp\left[\frac{i\hbar}{2}
\left(\lvec{\partial}_{q}\vec{\partial}_{p}-
\lvec{\partial}_{p}\vec{\partial}_{q}\right)\right]g.
\label{starMDef}
\end{equation}
This is obviously a special case of the circle product for the case where the symmetic part of $B$ vanishes.

We give a short review of the deformation quantization procedure
for the harmonic oscillator with Hamilton function
\begin{equation}
H(q,p)=\frac{p^2}{2m}+\frac{m\omega^2}{2}q^2. \label{harm}
\end{equation}
For more details see the review in \cite{HH}. The
Wigner functions $\pi_n^{(M)}$ and the energy levels $E_n$ of the
harmonic oscillator can be calculated with the help of the star
exponential
\begin{equation}
\mathrm{Exp}_M(Ht)=
\sum_{n=0}^{\infty}\frac{1}{n!}\left(\frac{-it}{\hbar} \right)^n
H^{n\starM}= \sum_{n=0}^{\infty}\pi_n^{(M)}e^{-iE_nt/\hbar},
\label{ExpDef}
\end{equation}
where $H^{n\starM}=H\starM\cdots\starM H$ is the $n$-fold star product
of $H$. For the harmonic oscillator one obtains
$E_n=\hbar\omega\left(n+\frac{1}{2}\right)$ and
\begin{equation}
\pi_n^{(M)}= 2(-1)^n
e^{-2H/\hbar\omega}L_n\left(\frac{4H}{\hbar\omega}\right),
\label{piMDef}
\end{equation}
where $L_n$ are the Laguerre polynomials. The energy levels and
the Wigner functions fulfill the $*$-genvalue equation
\begin{equation}
H\starM\pi_n^{(M)}= E_n\pi_n^{(M)}.
\end{equation}
The Wigner functions $\pi_n^{(M)}$ are normalized according to
\begin{equation} \frac {1}{2\pi\hbar}\int\pi_n^{(M)}\,dq\,dp=1.
\end{equation}
 The expectation value of a phase space function $f$ can be calculated
as
\begin{equation}
\langle f\rangle=\frac{1}{2\pi\hbar}\int
f\starM\pi_n^{(M)}\,dq\,dp.
\end{equation}

We now consider dynamical systems involving fermionic degrees of
freedom. These degrees of freedom are described by Grassmann
variables, so for one-dimensional systems no quadratic kinetic or
potential terms exist, because of the nilpotency of these
variables. The simplest non-trivial system in Grassmannian
mechanics is therefore a two dimensional system with Lagrange
function \cite{Ber}
\begin{equation}
L=\frac{i}{2}\left(\theta_1\dot{\theta}_1+\theta_2\dot{\theta}_2\right)
+i\omega\theta_1\theta_2,
\end{equation}
where $\theta_1,\theta_2$ are Grassmann variables. The canonical
momenta are
\begin{equation}
\rho_\alpha= -\frac{i}{2}\theta_{\alpha},\qquad \alpha=1,2,
\label{canomo}
\end{equation}
and the Hamilton function is given by
\begin{equation}
H= \dot{\theta}^{\alpha}\rho_\alpha-L= -i\omega\theta_1\theta_2.
\label{HfermDef}
\end{equation}
Eq.\ (\ref{canomo}) implies that this Hamiltonian may be seen as
describing rotation. Indeed, the fermionic angular momentum, which
corresponds to the spin, is
\begin{equation}
S_3=\theta_1\rho_2-\theta_2\rho_1= -i\theta_1\theta_2,
\label{Sdef}
\end{equation}
so that the Hamiltonian in (\ref{HfermDef}) can also be written as
$H=\omega S_3$. As a vector the angular momentum points out of the
$\theta_1$-$\theta_2$ plane. Therefore it is natural to consider
the two dimensional fermionic oscillator as embedded into a three
dimensional fermionic space with coordinates $\theta_1$,
$\theta_2$ and $\theta_3$. We choose units such that both the
fermionic coordinates and momenta have dimension $\sqrt{\hbar}$.

The appropriate star product for the quantization of fermionic
degrees of freedom is given by specifying $B(\theta_i,\theta_j)
=\frac{\hbar}{2}\delta_{ij}$ in the circle product
(\ref{circdef}), i.e.\
\begin{equation} F\starP G=
F\exp\left[\frac{\hbar}{2}\sum_{i=1}^d
\lvec{\partial}_{\theta_i}\vec{\partial}_{\theta_i}\right] G.
\label{starPDef}
\end{equation}
We call this product the Pauli star product, it is first mentioned
in \cite{Bayen}. It was shown in \cite{ferm} that it can be
obtained by deformation quantization of a Grassmann algebra. The
Pauli star product (\ref{starPDef}) leads to a Cliffordization of
the Grassmann algebra of the $\theta_i$, because the
star-anticommutator is given by
\begin{equation}
\{\theta_i,\theta_j\}_{\starP} =
\theta_i\starP\theta_j+\theta_j\starP\theta_i= \hbar\delta_{ij}.
\end{equation}
The even Grassmann functions
\begin{equation}
\sigma^i= \frac{1}{i\hbar}\varepsilon^{ijk}\theta_j\theta_k ,
\qquad i=1,2,3, \label{sigma}
\end{equation}
fulfill the relations
\begin{equation}
\left[\sigma^i,\sigma^j\right]_{\starP}=
2i\varepsilon^{ijk}\sigma^k \qquad\mathrm{and}\qquad
\left\{\sigma^i,\sigma^j\right\}_{\starP}= 2\delta^{ij}
\end{equation}
with $\left[\sigma^i,\sigma^j\right]_{\starP}=
\sigma^i\starP\sigma^j-\sigma^j\starP\sigma^i$, they therefore
correspond to the Pauli matrices.  Note that
$\{1,\sigma^1,\sigma^2,\sigma^3\}$ is a basis of the even
subalgebra of the Grassmann algebra, and that this space is closed
under $\starP$ multiplication. From Eqs.\ (\ref{Sdef}) and
(\ref{sigma}) we see that $S_3=\frac{\hbar}{2}\sigma^3$ and
$H=\omega S_3=\frac{\hbar\omega}{2}\sigma^3$.

The \emph{involution} operation in the space of Grassmann
variables \cite{Ber} is a mapping $F \mapsto \overline{F}$
satisfying the conditions
\begin{equation}
\overline{\overline{F}} = F \text{,} \qquad \overline{F_1 F_2} =
\overline{F_2}\, \overline{F_1} \qquad \text{and}\qquad
\overline{c F} = \bar c \overline{F}\text{,} \label{involution}
\end{equation}
where $c$ is a complex number and $\bar c$ its complex conjugate.
For the generators $\theta_i$ of the Grassmann algebra we assume
$\overline{\theta_i} = \theta_i$, so that for $\sigma_i$ defined
in (\ref{sigma}) the relation $\overline{\sigma_i} = \sigma_i$
holds true. This corresponds to the fact that the $2\times 2$
Pauli matrices $\hat{\sigma}^i$ are \emph{hermitian}.

The \emph{Hodge dual} maps a Grassmann monomial of grade $r$ into
a monomial of grade $d-r$, where $d$ is the number of Grassmann
basis elements:
\begin{equation}
\star\left(\theta_{i_1}\theta_{i_2}\cdots\theta_{i_r}\right) =
\frac{1}{(d-r)!}\varepsilon^{i_{r+1}\cdots i_d}_{i_1\cdots i_r}
\theta_{i_{r+1}}\cdots\theta_{i_d}. \label{hodge}
\end{equation}
With the help of the Hodge dual we can define the trace for $d=3$
as
\begin{equation}
\mathrm{Tr}(F)= \frac{2}{\hbar^3}\int
d\theta_3d\theta_2d\theta_1\,\star F. \label{TrDef}
\end{equation}
The integration is given by the Berezin integral, for which we have
$\int d\theta_i\,\theta_j=\hbar\delta_{ij}$, where $\hbar$ on the
right hand side is due to the fact that the variables $\theta_i$
have units of $\sqrt{\hbar}$. The only monomial with a non-zero
trace is $1$, so that by the linearity of the integral we obtain
the trace rules
\begin{equation}
\mathrm{Tr}(\sigma^i)= 0\qquad\mathrm{and}\qquad
\mathrm{Tr}(\sigma^i\starP\sigma^j)= 2\delta^{ij} \text{.}
\label{sigma-traces}
\end{equation}

With the fermionic star product (\ref{starPDef}) one can---as in
the bosonic case---calculate the energy levels and the
$*$-eigenfunctions of the fermionic oscillator \cite{ferm}. This
can be done by using the fermionic star exponential
\begin{equation}
\mathrm{Exp}_P(Ht)=
\sum_{n=0}^{\infty}\frac{1}{n!}\left(-\frac{it}{\hbar} \right)^n
H^{n\starP} = \pi_{1/2}^{(P)}e^{-i\omega
t/2}+\pi_{-1/2}^{(P)}e^{i\omega t/2}, \end{equation} where the
Wigner functions are given by
\begin{equation}
\pi_{\pm 1/2}^{(P)}= \frac{1}{2}\mp
\frac{i}{\hbar}\theta_1\theta_2 =
\frac{1}{2}\left(1\pm\sigma^3\right) \text{.} \label{piPDef}
\end{equation}
The $\pi_{\pm1/2}^{(P)}$ fulfill the $*$-genvalue equation
\begin{equation}
H\starP\pi_{\pm1/2}^{(P)}=E_{\pm1/2}\pi_{\pm1/2}^{(P)}
\end{equation}
for the energy levels
$E_{\pm1/2}=\pm\frac{\hbar\omega}{2}$. The Wigner functions
$\pi_{\pm1/2}^{(P)}$ are complete, idempotent and normalized with
respect to the trace, i.\,e.\ they fulfill the equations
\begin{equation}
\pi_{+1/2}^{(P)} +\pi_{-1/2}^{(P)} = 1, \qquad
\pi_{\alpha}^{(P)}\starP \pi_{\beta}^{(P)} =
\delta_{\alpha\beta}\pi_{\alpha}^{(P)} \qquad \text{and} \qquad
\mathrm{Tr} (\pi_{\pm1/2}^{(P)}) = 1,
\end{equation}
respectively. Furthermore they correspond to spin up and spin down
states since the Wigner functions (\ref{piPDef}) correspond to the
spin projectors, and the expectation values of the spin components
$\fett{S}=\frac{\hbar}{2}\fett{\sigma}$ are
\begin{subequations}
\begin{eqnarray}
\langle S_i\rangle&=&\mathrm{Tr}\left(\pi_{\pm1/2}^{(M)}\starP
\frac{\hbar}{2}\sigma^i\right)=0 \qquad (i=1,2),\\
\langle S_3\rangle&=&\mathrm{Tr}\left(\pi_{\pm1/2}^{(M)}\starP
\frac{\hbar}{2}\sigma^3\right) =\pm\frac{\hbar}{2},
\\\langle\fett{S}^{2\starP}\rangle&=&\mathrm{Tr}\left(
\pi_{\pm1/2}^{(M)}\starP\frac{\hbar^2}{4}\fett{\sigma}^{2\starP}
\right)=\frac{3}{4}\hbar^2\text{.}
\end{eqnarray}
\end{subequations}

The star exponential (\ref{ExpDef}) allows us to calculate the
time development of the $\sigma^i$ as
\begin{subequations}
\begin{eqnarray}
\sigma^1(t)&=&\mathrm{Exp}_P(-Ht)\starP\sigma^1\starP\mathrm{Exp}_P(Ht)
\,=\,
\sigma^1\cos(\omega t)-\sigma^2\sin( \omega t),\\
\sigma^2(t)&=&\mathrm{Exp}_P(-Ht)\starP\sigma^2\starP\mathrm{Exp}_P(Ht)
\,=\,
\sigma^1\sin(\omega t)+\sigma^2\cos(\omega t),\\
\sigma^3(t)&=&\mathrm{Exp}_P(-Ht)\starP\sigma^3\starP\mathrm{Exp}_P(Ht)
\,=\, \sigma^3.
\end{eqnarray}
\end{subequations}
With these expressions it is easy to see that the $*$-Heisenberg
equation
\begin{equation}
i\hbar\frac{df(t)}{dt}=\left[f(t),H(t)\right]_{\starP}
\end{equation}
for the spin is given by
\begin{equation}
\frac{dS_1(t)}{dt} = -\omega S_2(t),\qquad \frac{dS_2(t)}{dt}=
\omega S_1(t) \qquad\text{and}\qquad \frac{dS_3(t)}{dt}=0.
\end{equation}
For $\omega=\left(\frac{e}{mc}\right)B_3$, where $B_3$ is the
third component of the magnetic field
$\mbox{\boldmath$B$}=(0,0,B_3)$, this leads to the equation of
motion for the spin in a magnetic field:
\begin{equation}
\frac{d\mbox{\boldmath$S$}}{dt}=
\frac{e}{mc}\mbox{\boldmath$B$}\times \mbox{\boldmath$S$}.
\end{equation}

In the fermionic $\theta$-space the spin
$\fett{S}=\frac{\hbar}{2}\fett{\sigma}$ is the generator of
rotations, which are described by the star exponential
\begin{equation}
\mathrm{Exp}_P(\fett{\varphi} \cdot \fett{S})=
\cos\frac{\varphi}{2}-i(\fett{\sigma} \cdot \fett{n})
\sin\frac{\varphi}{2}, \label{RotStarExp}
\end{equation}
where $\fett{\varphi}=\varphi \fett{n}$ with the angle of rotation
$\varphi$ and rotation axis $\mbox{\boldmath $n$}$. The vector
$\fett{\theta}=(\theta_1,\theta_2,\theta_3)^T$ transforms
according to
\begin{equation}
\mathrm{Exp}_P(\fett{\varphi} \cdot \fett{S}) \starP
\fett{\theta}\starP \overline{\mathrm{Exp}_P(\fett{\varphi} \cdot
\fett{S})} = \mathrm{Exp}_P(\fett{\varphi} \cdot \fett{S}) \starP
\fett{\theta}\starP \mathrm{Exp}_P(-\fett{\varphi} \cdot \fett{S}) =
R(\fett{\varphi}) \fett{\theta} \label{rotation}
\end{equation}
where $R(\fett{\varphi})$ is the rotation matrix which satisfies
\begin{equation}
R(\fett{\varphi})\fett{\theta}=\fett{n}(\fett{n}\cdot\fett{\theta})+\cos\varphi(\fett{\theta}
-\fett{n}(\fett{n}\cdot\fett{\theta}))-\sin \varphi(\fett{n}\times
\fett{\theta}).
\end{equation}
The axial vector $\fett{\sigma}$ transforms in the same way under
rotations.

\section{Charged Particle with Spin in a Constant Magnetic Field}
\setcounter{equation}{0} \label{mag}

\quad The bosonic and the fermionic oscillators can be combined to
treat a physical system consisting of a charged particle with spin
in a constant magnetic field in the star product formalism. We
first consider the bosonic part of this problem: a charged
spinless particle in a constant magnetic field. The magnetic field
points in the direction of $q_3$ and can be described with the
gauge potential $\mbox{\boldmath$A$}=\frac{B}{2}(-q_2,q_1,0)$. By
minimal substitution
$\mbox{\boldmath$p$}\rightarrow\mbox{\boldmath$p$}
-\frac{e}{c}\mbox{\boldmath$A$}$ one obtains the Landau
Hamiltonian
\begin{equation}
H_L= \frac{1}{2m}\left( \tilde{p}_1^2+\tilde{p}_2^2\right),
\end{equation}
where we have defined
\begin{equation}
\tilde{p}_1= p_1-\frac{e}{c}A_1= p_1+\frac{m\omega}{2}q_2
\qquad\mathrm{and}\qquad \tilde{p}_2= p_2-\frac{e}{c}A_2=
p_2-\frac{m\omega}{2}q_1
\end{equation}
with $\omega=\frac{eB}{mc}$. In order to quantize this two
dimensional system we transform the Moyal product (\ref{starMDef})
into the $(q_i,\tilde{p}_i)$-coordinates; the resulting expression
is
\begin{equation}
f \tstarM g= f\exp\left[\frac{i\hbar}{2}
\left(\lvec{\partial}_{q_1}\vec{\partial}_{\tilde{p}_1}
-\lvec{\partial}_{\tilde{p}_1}\vec{\partial}_{q_1}
+\lvec{\partial}_{q_2}\vec{\partial}_{\tilde{p}_2}
-\lvec{\partial}_{\tilde{p}_2}\vec{\partial}_{q_2}\right)
+\frac{i\hbar m\omega}{2}
\left(\lvec{\partial}_{\tilde{p}_1}\vec{\partial}_{\tilde{p}_2}
-\lvec{\partial}_{\tilde{p}_2}\vec{\partial}_{\tilde{p}_1}\right)\right]g
. \label{tstarMDef}
\end{equation}

The $*$-genvalue equation
\begin{equation}
\frac{1}{2m}\left(\tilde{p}_1^2+\tilde{p}_2^2
\right)\,\tstarM\pi_n^{(\tilde{M})}=E_n\pi_n^{(\tilde{M})}
\end{equation}
can easily be solved by comparison with the bosonic oscillator. As
we have seen above the $*$-eigenfunctions of the bosonic
oscillator depend only on the Hamiltonian. Therefore we also
expect $\pi_n^{(\tilde{M})}$ to depend on $\tilde{p}_1$ and
$\tilde{p}_2$ only. Taking this as an ansatz, only the second part
of the star product (\ref{tstarMDef}), which can be written as
\begin{equation}
\exp\left[\frac{i\hbar}{2}\left(
\lvec{\partial}_{\left(\frac{\tilde{p}_1}{m\omega}\right)}
\vec{\partial}_{\tilde{p}_2} -\lvec{\partial}_{\tilde{p}_2}
\vec{\partial}_{\left(\frac{\tilde{p}_1}{m\omega}\right)}
\right)\right], \label{secpa}
\end{equation}
has to be taken into account for the $*$-genvalue equation.

Setting $q=\frac{\tilde{p}_1}{m\omega}$ and $p=\tilde{p}_2$ the
Landau Hamiltonian $H_L$ reduces to the Hamiltonian of the bosonic
harmonic oscillator (\ref{harm}) and Eq.~(\ref{secpa}) becomes the
Moyal product in canonical variables. Then it is clear that the
$*$-eigenfunctions of the Landau Hamiltonian are in analogy to
(\ref{piMDef}) given by
\begin{equation}
\pi_n^{(\tilde{M})} (\tilde{p}_1,\tilde{p}_2)=
\pi_n^{(\tilde{M})}(H_L) =
2(-1)^n\exp\left(-\frac{2H_L}{\hbar\omega}\right)L_n
\left(\frac{4H_L}{\hbar\omega}\right)\text{.}
\end{equation}
The energy levels are the Landau levels
$E_n=\hbar\omega\left(n+\frac{1}{2}\right)$.

Since the system considered here is described in a four
dimensional phase space we
expect that another observable which commutes with the
Hamiltonian is needed to characterize all the energy
$*$-genfunctions. To find such an observable it is useful to write
the star product (\ref{tstarMDef}) in the two forms
\begin{subequations}
\begin{eqnarray}
f \tstarM g &=& f \exp\left[\frac{i\hbar}{2}
\left(\lvec{\partial}_{q_1}\vec{\partial}_{\tilde{p}_1}
-\lvec{\partial}_{\tilde{p}_1} \big(\vec{\partial}_{q_1} - m\omega
\vec{\partial}_{\tilde{p}_2}\big)
+\lvec{\partial}_{q_2}\vec{\partial}_{\tilde{p}_2}
-\lvec{\partial}_{\tilde{p}_2}\big(\vec{\partial}_{q_2} + m\omega
\vec{\partial}_{\tilde{p}_1}\big)\right) \right]g \\
& = & f\exp\left[\frac{i\hbar}{2} \left(\big(\lvec{\partial}_{q_1}
- m\omega \lvec{\partial}_{\tilde{p}_2}\big)
\vec{\partial}_{\tilde{p}_1}
-\lvec{\partial}_{\tilde{p}_1}\vec{\partial}_{q_1}
+\big(\lvec{\partial}_{q_2} + m\omega
\lvec{\partial}_{\tilde{p}_1}\big)\vec{\partial}_{\tilde{p}_2}
-\lvec{\partial}_{\tilde{p}_2}\vec{\partial}_{q_2}\right)\right]g
\label{tstarM2b}
\end{eqnarray}\label{tstarM2}\end{subequations}by simply rearranging the terms in the argument of the exponential
function. By observing that the functions $\tilde{q}_1 =
q_1+\frac{1}{m\omega}\tilde{p}_2$ and $\tilde{q}_2 =
q_2-\frac{1}{m\omega}\tilde{p}_1$ fulfill the equations
\begin{equation}
(\partial_{q_1}-m\omega\partial_{\tilde{p}_2})\tilde{q}_i= 0
\qquad\mathrm{and}\qquad
(\partial_{q_2}+m\omega\partial_{\tilde{p}_1})\tilde{q}_i= 0
\label{eigsch}
\end{equation}
it is obvious from Eqs.\ (\ref{tstarM2}) that every function of
the $\tilde{q}_i$ commutes with every function of the
$\tilde{p}_i$, e.\,g.
\begin{equation}
H_L \tstarM f(\tilde{q}_1,\tilde{q}_2)= H_L
f(\tilde{q}_1,\tilde{q}_2)= f(\tilde{q}_1,\tilde{q}_2) H_L=
f(\tilde{q}_1,\tilde{q}_2) \tstarM H_L \text{.} \label{starg}
\end{equation}
This implies that $\tilde{q}_1$ and $\tilde{q}_2$ are conserved
phase space functions, and also that all functions of the form
$f(\tilde{q}_1,\tilde{q}_2)\pi_n^{(\tilde{M})}(\tilde{p}_1,\tilde{p}_2)$
are $*$-genfunctions of the Hamiltonian. Obviously such a function
becomes a $*$-genfunc\-tion of the angular momentum
\begin{equation}
J= q_1p_2-q_2p_1= -\frac{1}{\omega}H_L+\frac{m\omega}{2}
(\tilde{q}_1^2+\tilde{q}_2^2) \label{JDef}
\end{equation}
if $f(\tilde{q}_1,\tilde{q}_2)$ is chosen to be a $*$-genfunction
of $\frac{m\omega}{2} (\tilde{q}_1^2+\tilde{q}_2^2)$.

Using Eq.\ (\ref{eigsch}) only two terms in the argument of the
exponential function contribute to the star product
(\ref{tstarM2b}) in the $*$-genvalue Eq.\ (\ref{starg}), so that
\begin{eqnarray}
\frac{m\omega^2}{2}(\tilde{q}_1^2+\tilde{q}_2^2)\ \tstarM
f(\tilde{q}_1,\tilde{q}_2)
&=&\frac{m\omega^2}{2}(\tilde{q}_1^2+\tilde{q}_2^2)
\exp\left[\frac{i\hbar}{2}\left(-\lvec{\partial}_{\tilde{p}_1}
\vec{\partial}_{q_1}-\lvec{\partial}_{\tilde{p}_2}\vec{\partial}_{q_2}\right)
\right]\nonumber\\
&=&\frac{m\omega^2}{2}(\tilde{q}_1^2+\tilde{q}_2^2)
\exp\left[\frac{i\hbar}{2m\omega}\left(\lvec{\partial}_{\tilde{q}_2}
\vec{\partial}_{\tilde{q}_1}-\lvec{\partial}_{\tilde{q}_1}
\vec{\partial}_{\tilde{q}_2}\right)\right]
f(\tilde{q}_1,\tilde{q}_2),
\end{eqnarray}
where we used the definition of $\tilde{q}_i$ in the last step.
Setting $q=\tilde{p}_2$ and $p=m\omega\tilde{q}_1$, the problem
reduces to the one dimensional harmonic oscillator,
so that $f(\tilde{q}_1,\tilde{q}_2)$ becomes
\begin{equation}
\pi_l^{(\tilde{M})}(\tilde{q}_1,\tilde{q}_2)=
2(-1)^l\exp\left(-\frac{m\omega}{\hbar}(\tilde{q}_1^2+\tilde{q}_2^2)\right)
L_l\left(\frac{2m\omega}{\hbar}(\tilde{q}_1^2+\tilde{q}_2^2)\right)
\end{equation}
and the $*$-eigenvalues of
$\frac{m\omega}{2}(\tilde{q}_1^2+\tilde{q}_2^2)$ are
$\hbar\left(l+\frac{1}{2}\right)$.

Thus, the Wigner functions of the Landau problem are
$\pi_{nl}^{(\tilde{M})}(\tilde{q}_1,\tilde{q}_2,\tilde{p}_1,\tilde{p}_2)$
$=\pi_{l}^{(\tilde{M})}(\tilde{q}_1,\tilde{q}_2)
\pi_{n}^{(\tilde{M})}(\tilde{p}_1,\tilde{p}_2)$ and lead with the
$*$-genvalue equation
\begin{equation} H_L
\starM\pi_{nl}^{(\tilde{M})}=E_n \pi_{nl}^{(\tilde{M})}
\end{equation}
to the Landau levels $E_n=\hbar\omega\left(n +\frac{1}{2}\right)$,
whereas the equation $J\tstarM\pi_{nl}^{(\tilde{M})}
=j_{nl}\pi_{nl}^{(\tilde{M})}$ gives rise to the angular momentum
eigenvalues $j_{nl}=\hbar(l-n)$. For a treatment of the problem of
a charged particle in a constant magnetic field using star
products and holomorphic coordinates see Ref.\ \cite{Landau}.

In order to include the interaction of the spin with the magnetic
field one has to consider both the bosonic and the fermionic
sectors. Therefore we first combine the bosonic and the fermionic
star products to the Moyal Pauli star product
\begin{equation}
F\starMP G= F\exp\left[\frac{i\hbar}{2}\sum_{i=1}^3
\left(\lvec{\partial}_{q_i}\vec{\partial}_{p_i}-
\lvec{\partial}_{p_i}\vec{\partial}_{q_i}\right)
+\frac{\hbar}{2}\sum_{i=1}^3\lvec{\partial}_{\theta_i}
\vec{\partial}_{\theta_i}\right]G. \label{starMPDef}
\end{equation}
The Poisson bracket corresponding to this star product was
considered in \cite{Soroka}. The realization of the ``Feynman
Trick'' \cite{Sakurai} for including the influence of a magnetic
field is given in the star product formalism by \begin{equation}
\left[\left(\mbox{\boldmath$p$}-\frac{e}{c}\mbox{\boldmath$A$}\right)\cdot
\mbox{\boldmath$\sigma$}\right]^{2\starMP} =
\left(\mbox{\boldmath$p$}-\frac{e}{c}\mbox{\boldmath$A$}\right)^{2\starM}
-\frac{\hbar
e}{c}\mbox{\boldmath$\sigma$}\cdot\mbox{\boldmath$B$}.
\end{equation}
Therefore an interaction term $H_I=-\frac{\hbar
e}{2mc}\mbox{\boldmath$\sigma$}\cdot\mbox{\boldmath$B$}$ is
induced in the Hamiltonian. With Eq.\ (\ref{sigma}) the
interaction term can be written as
\begin{equation}
H_I= -\frac{e\hbar}{2mc}B\sigma^3= -i\omega\theta_1\theta_2,
\end{equation}
which is just the fermionic oscillator (\ref{HfermDef}). Thus, the
system describing a charged particle with spin in a constant
magnetic field can be regarded as the sum of a two dimensional
bosonic and fermionic oscillators with the same parameter
$\omega$, i.e.\ {\em a supersymmetric oscillator} \cite{Witten}.
The projectors for this system are products of the projectors of
the bosonic and the fermionic oscillators, i.e.\
$\pi_{n}^{(\tilde{M})}\pi_{\pm1/2}^{(P)}$ and the energy levels
are the sum of the single energy levels: $E_{n,\pm1/2}
=\hbar\omega\left(n+\frac{1}{2}\pm\frac{1}{2}\right)$.

\section{The Supersymmetric Oscillator and the Witten Index}
\setcounter{equation}{0} \label{susy}

\qquad The harmonic oscillator can be factorized into
$H=\omega\bar{a}a$ by using the holomorphic coordinates
\begin{equation}
a=\sqrt{\frac{m\omega}{2}}\left(q+i\frac{p}{m\omega}\right)\qquad
\text{and}\qquad
\bar{a}=\sqrt{\frac{m\omega}{2}}\left(q-i\frac{p}{m\omega}\right).
\end{equation}
The Moyal product for holomorphic coordinates is
\begin{equation}
f\starM g= f\exp\left[\frac{\hbar}{2}\left(
\lvec{\partial}_a\vec{\partial}_{\bar{a}}
-\lvec{\partial}_{\bar{a}}\vec{\partial}_{a}\right)\right]g.
\label{starholobo}
\end{equation}
Since the projectors $\pi_n^{(M)}$ depend on $H$ only, they have
the same functional form as in Eq.\ (\ref{piMDef}).

In the fermionic case one can also go over to holomorphic
coordinates
\begin{equation}
f=\frac{1}{\sqrt{2}}(\theta_2+i\theta_1)\qquad {\rm and}\qquad
\bar{f}=\frac{1}{\sqrt{2}} (\theta_2-i\theta_1).
\end{equation}
The fermionic oscillator in these coordinates has the form
$H=\omega\bar{f}f$ and the Pauli star product becomes
\begin{equation}
F\starP G= F\exp\left[\frac{\hbar}{2}\left(
\lvec{\partial}_f\vec{\partial}_{\bar{f}}
+\lvec{\partial}_{\bar{f}}\vec{\partial}_{f}\right)\right]G.
\label{starholoferm}
\end{equation}
The fermionic projectors in holomorphic coordinates are
$\pi_{\pm1/2}^{(P)} =\frac{1}{2}\pm\frac{1}{\hbar}\bar{f}f$.

The bosonic and the fermionic oscillator Hamiltonian can be
combined to the supersymmetric Hamiltonian
$H=\omega(\bar{a}a+\bar{f}f)$. The corresponding supersymmetric
star product consists of the bosonic and fermionic star products
(\ref{starholobo}) and (\ref{starholoferm}):
\begin{equation}
F\starSU G= F\exp\left[\frac{\hbar}{2}\left(
\lvec{\partial}_a\vec{\partial}_{\bar{a}}
-\lvec{\partial}_{\bar{a}}\vec{\partial}_{a}
+\lvec{\partial}_f\vec{\partial}_{\bar{f}}
+\lvec{\partial}_{\bar{f}}\vec{\partial}_{f}\right)\right]G.
\end{equation}
The supersymmetric projectors
$\pi_{n_F,n_B}^{(SU)}=\pi_{n_B}^{(M)} \pi_{n_F=\pm1/2}^{(P)}$ are
products of the bosonic and the fermionic projectors.

One can define the functions $Q_+ =
\frac{1}{\sqrt{\hbar}}a\bar{f}$ and $Q_-=
\frac{1}{\sqrt{\hbar}}\bar{a}f$ , which satisfy
\begin{equation}
Q_+\starSU\pi_{n_F,n_B}^{(SU)}\starSU Q_-=
\hbar\pi_{n_F+1,n_B-1}^{(SU)} \qquad\text{and}\qquad
Q_-\starSU\pi_{n_F,n_B}^{(SU)}\starSU Q_+=
\hbar\pi_{n_F-1,n_B+1}^{(SU)},
\end{equation}
and thus relate the otherwise distinct bosonic and fermionic
sectors. The functions $\pi_{\pm1/2}^{(P)}$, $Q_+$ and $Q_-$
fulfill the relations
\begin{equation}
\pi_{\pm1/2}^{(P)}\starSU\pi_{\pm1/2}^{(P)}= \pi_{\pm1/2}^{(P)}
,\qquad Q_{\pm}\starSU\pi_{\mp1/2}^{(P)}= Q_{\pm}
\qquad\text{and}\qquad \pi_{\pm1/2}^{(P)}\starSU Q_{\pm}= Q_{\pm},
\end{equation}
so that these functions form a Fredholm quadruple $\Xi$, with
which one can define the index \cite{Fedosov}
\begin{eqnarray}
\mathrm{ind}\Xi
&=&\mathrm{tr}\left[\pi_{-1/2}^{(P)}-\frac{1}{\hbar}Q_+\starSU
Q_-\right]
-\mathrm{tr}\left[\pi_{+1/2}^{(P)}-\frac{1}{\hbar}Q_-\starSU
Q_+\right]
\nonumber\\
&=&\mathrm{tr}\left[\pi_{-1/2}^{(P)}
\left(\frac{1}{2}-\frac{a\bar{a}}{\hbar}\right)\right]
-\mathrm{tr}\left[\pi_{+1/2}^{(P)}
\left(\frac{1}{2}-\frac{a\bar{a}}{\hbar}\right)\right],
\label{indXi}
\end{eqnarray}
where the trace \textquotedblleft$\mathrm{tr}$\textquotedblright\
is the sum over all states
\begin{equation}
\mathrm{tr}[F]= \sum_{n=0}^{\infty}\sum_{n_s=\pm1/2}\int
d^2a\,\mathrm{Tr} (\pi_n^{(M)}\pi_{n_s}^{(P)}\starSU F).
\end{equation}
The trace \textquotedblleft$\mathrm{Tr}$\textquotedblright\ is
defined as in (\ref{TrDef}). The second terms in the round
brackets of (\ref{indXi}) give the sum of the numbers of the
bosonic states. Since all bosonic states with $E>0$ appear as
pairs in the bosonic and the fermionic sector, these two terms
cancel each other. The first term in the round brackets counts the
number of bosonic states, so that the index is the difference of
the number of bosonic states in the bosonic and the fermionic
sector. Because of the pairing of states with $E>0$ the index will
be zero if there is a state with $E=0$ in the bosonic and the
fermionic sector and one if only one of the sectors has a state
with $E=0$. This index is the Witten index \cite{Wittenindex},
which reveals whether the supersymmetry is exact or broken.

\section{The Dirac Equation}
\setcounter{equation}{0}\label{rqm}

\qquad With the Grassmannian representation of the Pauli matrices
(\ref{sigma}) it is possible to give a Grassmannian representation
of the Dirac $\gamma$-matrices with two sets of $\sigma^i$. Starting
with the variables $\theta_1,\ldots,\theta_6$ one can build two
triples of $\sigma^i= \frac{2}{i\hbar} \varepsilon^{ijk}
\theta_j\theta_k$, one for $i,j,k\in\{1,2,3\}$ and one for
$i,j,k\in\{4,5,6\}$, by which the tensor structure of
$\hat{\alpha}^i=\hat{\sigma}^1\otimes\hat{\sigma}^i$ and
$\hat{\beta}=\hat{\sigma}^3\otimes \hat{I}$ in the Dirac
representation as $4\times 4$ matrices is reproduced. The four
functions defined as
\begin{equation}
\alpha^i=\sigma^i\sigma^4\quad (i=1,2,3)
\qquad\mathrm{and}\qquad\beta=\sigma^6, \label{albesix}
\end{equation}
fulfill the equations
\begin{equation}
\{\alpha_k,\alpha_l\}_{\starP}= 2\delta_{kl}\ ,\qquad
\{\alpha_k,\beta\}_{\starP}= 0\qquad\mathrm{and}\qquad
\beta\starP\beta= 1\text{,} \label{Diral}
\end{equation}
where we used the Pauli star product (\ref{starPDef}) for $d=6$.
Conceptually we have turned around Dirac's ansatz. While Dirac
tried to find (matrix) quantities $\hat{\alpha}$ and $\hat{\beta}$
that fulfill the Dirac algebra, we look for a product such that
the relations of the Dirac algebra are fulfilled. This leads us to
the Pauli star product.

In this approach to the Dirac theory we combined two copies of the
three dimensional fermionic spaces which in Sec.~5 appeared to be
suitable to describe spin. Thereby the subalgebra of the Grassmann
algebra which contains only elements of even grade was used. From
the algebraic point of view one can ask whether it is necessary to
use a Grassmann algebra with six generators to reproduce the Dirac
algebra (\ref{Diral}). Indeed, the functions
\begin{equation}
\alpha^i = \sqrt{\frac{2}{\hbar}}\sigma^i\theta_5
\qquad\text{and}\qquad \beta= \frac{2i}{\hbar}\theta_4\theta_5
\label{albefive}
\end{equation}
also fulfill the Dirac algebra (\ref{Diral}) by using five
Grassmann variables and the star product (\ref{starPDef}) for
$d=5$. One is lead to this representation by constructing the
Dirac Hamiltonian as a supercharge from supersymmetric quantum
mechanics \cite{ferm}.

Since the Clifford algebra of the Dirac matrices is four
dimensional, it should also be possible to start with a Grassmann
algebra generated by $\theta_1,\ldots,\theta_4$ that is turned
into a Clifford algebra with the Pauli star product
(\ref{starPDef}) for $d=4$. Indeed the dimensionless variables
\begin{equation}
\alpha^i= \sqrt{\frac{2}{\hbar}}\theta_i \qquad \text{and}\qquad
\beta= \sqrt{\frac{2}{\hbar}}\theta_4 \label{albefour}
\end{equation}
obey the relations (\ref{Diral}) and form another representation
of the Dirac algebra. With respect to the Pauli star product the
generators of the Grassmann algebra become here generators of the
Clifford algebra, as in the Cliffordization procedure
described in the first sections.

The four dimensional representation of the Dirac algebra can also
be motivated by considerations of the symmetries of spacetime.
With the definition of $\sigma^i$ in Eq.\ (\ref{sigma}) we can
reproduce the commutation relations of the corresponding Pauli
matrices, and in Eq.\ (\ref{rotation}) it was shown that
$S_i=\frac{\hbar}{2} \sigma^i$ generate rotations in the Grassmann
algebra. So far only the even part of the Grassmann algebra was
involved, so that the question arises what transformations the
$\theta_i$ are related to. The definition
$K_i=i\sqrt{\hbar/2}\,\theta_i=i \frac{\hbar}{2}\, \alpha^i$ leads
to the commutation relations
\begin{equation}
\left[S_i,S_j\right]_{\starP}= i\hbar\varepsilon^{ijk}S^k
\text{,}\qquad \left[S_i,K_j\right]_{\starP}=
i\hbar\varepsilon^{ijk}K_k \qquad\text{and}\qquad
\left[K_i,K_j\right]_{\starP}= -i\hbar\varepsilon^{ijk}S_k\text{,}
\end{equation}
so that we can identify the components of $\fett{K}$ as generators of the Lorentz
boosts. The star exponential
$\mathrm{Exp}_P(\fett{\omega}\cdot\fett{K})$ transforms
$\alpha^\mu = (1,\fett{\alpha})$ with $\mu=0,1,2,3$ like a four
vector:
\begin{equation}
\mathrm{Exp}_P\left(\fett{\omega}\cdot\fett{K}\right) \starP
\alpha^\mu \starP
\overline{\mathrm{Exp}_P\left(\fett{\omega}\cdot\fett{K}\right)} =
\mathrm{Exp}_P\left(\fett{\omega}\cdot\fett{K}\right) \starP
\alpha^\mu \starP
\mathrm{Exp}_P\left(\fett{\omega}\cdot\fett{K}\right) =
\Lambda^\mu_{\ \nu}(\fett{\omega})\, \alpha^\nu \label{boost1}
\text{,}
\end{equation}
where $\Lambda^\mu_{\ \nu}(\fett{\omega})$ is the matrix
representation of a Lorentz boost. In contrast to Eqs.\
(\ref{rotation}) the signs of the parameters---$\fett{\omega}$ in
this case---are not changed by the involution because
$\overline{\fett{K}}=-\fett{K}$ compared to
$\overline{\fett{S}}=\fett{S}$.

As one can see in Eq.\ (\ref{rotation})
$\fett{\alpha}=\sqrt{2/\hbar}\,\fett{\theta}$ behaves like a
vector under rotation and therefore should be mapped into
$\mathcal{P}(\fett{\alpha})=-\fett{\alpha}$ by the parity
transformation $\mathcal{P}$, which cannot be represented without
a further extension of the algebra. By introducing an additional
generator $\theta_4$ to the three dimensional Grassmann algebra
and extending the star exponential (\ref{starPDef}) to $d=4$ a
representation of the parity transformation can be given by
\begin{equation}
\mathcal{P}(F)=\beta\starP F \starP \beta \label{par}
\end{equation}
with the definition $\beta = \sqrt{2/\hbar}\,\theta_4$. The scalar
$1$ and the axial vector $\fett{\sigma}$ defined in (\ref{sigma})
are indeed invariant with respect to this transformation.

The three representations (\ref{albesix}), (\ref{albefive}) and
(\ref{albefour}) were built by starting with a representation of
rotations in a Grassmann algebra with generators
$\{\theta_1,\theta_2,\theta_3\}$, and as such the rotations are
generated by
\begin{equation}
S_i = \frac{\hbar}{2}\,\sigma^i = \frac{1}{2i}
\varepsilon^{ijk} \theta_j \theta_k = -i\frac{\hbar}{4}
\varepsilon^{ijk}\alpha^j\starP\alpha^k
\end{equation}
in all representations. Since $S_i$ can be given solely in terms
of $\alpha^i$ and since all versions of the Dirac algebra
$\{\alpha^i,\beta\}$ are equivalent, $\fett{\alpha}$ behaves like
a vector in all three representations, i.\,e.\
\begin{equation}
\mathrm{Exp}(\fett{\varphi} \cdot \fett{S}) \starP
\fett{\alpha}\starP \mathrm{Exp}(-\fett{\varphi} \cdot \fett{S}) =
R(\fett{\varphi}) \fett{\alpha} \text{.} \label{rotation2}
\end{equation}
The same argumentation is also valid for Lorentz boosts generated
by $K_i=i\frac{\hbar}{2}\,\alpha^i$ and the parity transformation
with $\beta$---thus equations (\ref{boost1}) and (\ref{par}) hold
true for all three definitions of $\{\alpha^i,\beta\}$.

It will now be shown that the rotations (\ref{rotation2}) and the
Lorentz boosts (\ref{boost1}) can be combined into one equation.
Before doing so it is useful to introduce Grassmann functions that
correspond to the Dirac matrices~$\hat\gamma^\mu$:
\begin{equation}
\gamma^0= \beta \qquad \text{and}\qquad \gamma^i=
\beta\starP\alpha^i \qquad\Rightarrow\qquad
\{\gamma^\mu,\gamma^\nu\}_{\starP}= 2g^{\mu\nu} \text{.}
\label{gammadef}
\end{equation}
Eq.\ (\ref{boost1}) is multiplied with $\beta$ from the left in
order to get the different signs in the two star exponentials
which occur in Eq.\ (\ref{rotation2}) for the rotations. Since
$\beta$ anticommutates with $K_i\propto\alpha^i$, Eq.\
(\ref{boost1}) becomes
\begin{equation}
\mathrm{Exp}_P\left(-\fett{\omega}\cdot\fett{K}\right) \starP
\gamma^\mu \starP
\mathrm{Exp}_P\left(\fett{\omega}\cdot\fett{K}\right) =
\Lambda^\mu_{\ \nu}(\fett{\omega})\, \gamma^\nu \label{boost2}
\text{.}
\end{equation}
With the definition
$\sigma^{\mu\nu}=\frac{i}{2}[\gamma^\mu,\gamma^\nu]_{\starP}$ the six
generators of the Lorentz transformation can be written as
\begin{subequations}
\begin{eqnarray}
K_i &=& i\frac{\hbar}{2} \alpha^i \ =\ i\frac{\hbar}{2}
\gamma^0\starP\gamma^i
\ =\ \frac{\hbar}{2} \sigma^{0i}, \label{Ki} \\
S_i &=& -i \frac{\hbar}{4} \varepsilon^{ijk} \alpha^j\starP
\alpha^k \ =\ i \frac{\hbar}{4} \varepsilon^{ijk}
\gamma^i\starP\gamma^k \ =\ \frac{\hbar}{2} \sum_{j<k}
\varepsilon^{ijk} \sigma^{jk} \text{.} \label{Si}
\end{eqnarray}
\end{subequations}
Therefore all Lorentz transformations are generated by
$\frac{\hbar}{2}\sigma^{\mu\nu}$ with $\mu<\nu$. Because $\beta$
commutes with $S_i\propto
\varepsilon^{ijk}\alpha^j\starP\alpha^k$, one can replace
$\fett{\alpha}$ by $\fett{\gamma}$ in Eq.\ (\ref{rotation2}) and
the resulting equation can finally be unified with (\ref{boost2})
to
\begin{equation}
\mathrm{Exp}_P\left(-\frac{\hbar}{4}\sigma^{\mu\nu}\omega_{\mu\nu}\right)
\starP\gamma^{\mu}\starP
\mathrm{Exp}_P\left(+\frac{\hbar}{4}\sigma^{\mu\nu}\omega_{\mu\nu}\right)
= \Lambda^{\mu}_{\ \nu}(\omega_{\mu\nu})\,\gamma^{\nu}\text{.}
\label{LT}
\end{equation}
This is the usual form of Lorentz transformation known from Dirac
theory.

For all representations of the Clifford algebra with $d=4$, $5$ or
$6$ generators $\theta_i$ a trace can be defined as in Eq.\
(\ref{TrDef}):
\begin{equation}
\mathrm{Tr}(F)= \frac{4}{\hbar^d}\int d\theta_d d\theta_{d-1}
\ldots d\theta_2 d\theta_1 \,\star F \label{TrDef2}
\end{equation}
and with ${\rm Tr}(1)=4$ all the well-known trace rules for the
$\gamma$-matrices are reproduced. The trace $\mathrm{Tr}(F)$
projects out the part of $F$ that is proportional to 1 just like
the map $\varepsilon$ that was used in Eq.\ (\ref{theo}), which is
the fermionic version of taking the vacuum expectation value. So
$\varepsilon$ can be made explicit by a Berezin integral.

With $\alpha_i$ and $\beta$ the Dirac Hamiltonian is given by
\begin{equation}
H_D= c\,\fett{\alpha}\cdot \fett{p}+\beta mc^2 \label{HDDef}
\end{equation}
and by using $H_D\starMP H_D=c^2\fett{p}^2+m^2c^4$ one can
calculate the star exponential as
\begin{equation}
\mathrm{Exp}_{MP}(H_Dt)=
\sum_{n=0}^{\infty}\frac{1}{n!}\left(\frac{t}{i\hbar} \right)^n
H_D^{n\starMP} = \pi_{-E}^{(MP)}(\fett{p})\,e^{+itE/\hbar}
+\pi_{+E}^{(MP)}(\fett{p})\,e^{-itE/\hbar}
\end{equation}
with the Wigner functions
\begin{equation}
\pi_{\pm E}^{(MP)}(\fett{p})= \frac{1}{2}\left(1 \pm\frac{H_D}{E}
\right) \label{diracpmEpi}
\end{equation}
and $E=\sqrt{c^2\fett{p}^2+m^2 c^4}$. The energy projectors
$\pi_{\pm E}^{(MP)}(\fett{p})$ are idempotent, complete and
fulfill the \mbox{$*$-genvalue} equations
\begin{equation}
H_D\starMP\pi_{\pm E}^{(MP)}(\fett{p})= \pm E\,\pi_{\pm
E}^{(MP)}(\fett{p})\text{.} \label{ham}
\end{equation}

One can find projectors that are $*$-genfunctions of the spin as
well, which is defined by the equation $\Su =
\frac{\hbar}{2}\gamma^5 \starP(\fett{\gamma} \cdot \fett{u})$. The
quantization axis $\fett{u}$ is a unit vector orthogonal to
$\fett{p}$, so that the equations $\Su\starP
\Su=\left(\frac{\hbar}{2}\right)^2$ and $[H_D,\Su]_{*P}=0$ hold.
For $\Su$ the star exponential is
\begin{equation}
\mathrm{Exp}_P(\Su \varphi) =
\sum_{n=0}^{\infty}\frac{1}{n!}\left(\frac{\varphi}{i\hbar}
\right)^n \Su^{n\starP} =
\pi_{-s}^{(P)}(\fett{u})\,e^{+i\varphi/2}+\pi_{+s}^{(P)}(\fett{u})\,e^{-i\varphi/2}
\end{equation}
with the Wigner functions
\begin{equation}
\pi_{\pm s}^{(P)}(\fett{u})= \frac{1}{2}\pm \frac{1}{\hbar}\Su
\text{.}\label{Susol}
\end{equation}
These are the star product analogues of the Dirac spin projectors
and they obey the $*$-genvalue equation
\begin{equation}
\Su\starP\pi_{\pm s}^{(P)}(\fett{u}) = \pm\frac{\hbar}{2}\pi_{\pm
s}^{(P)}(\fett{u}) \text{.}\label{Sugl}
\end{equation}

Since we have for $\mbox{\boldmath$p$}\cdot\mbox{\boldmath$u$}=0$:
\begin{equation}
\left[\beta,\gamma^5\starP(\mbox{\boldmath$\gamma$}\cdot
\mbox{\boldmath$p$})\right]_{\starP}= 0 \qquad\text{and}\qquad
\left[\mbox{\boldmath$p$}\cdot\mbox{\boldmath$\alpha$},
\gamma^5\starP(\mbox{\boldmath$\gamma$}
\cdot\mbox{\boldmath$u$})\right]_{\starP}= 0
\label{koms}
\end{equation}
the Wigner functions $\pi_{\pm E}^{(MP)}(\fett{p})$ and $\pi_{\pm
s}^{(P)}(\fett{u})$ and the observables $H_D$ and $\Su$ commute
under the star product. The Wigner functions for the Dirac problem
are therefore given by
\begin{equation}
\pi_{\pm E,\pm s}^{(MP)}(\fett{p},\fett{u})= \pi_{\pm
E}^{(MP)}(\fett{p}) \starMP \pi_{\pm s}^{(P)}(\fett{u})
\label{pidi}
\end{equation}
and the $*$-genvalue equations are
\begin{equation}
H_D\starMP\pi_{\pm E,\pm s}^{(MP)}(\fett{p},\fett{u})= \pm
E\pi_{\pm E,\pm s}^{(MP)}(\fett{p},\fett{u})
\qquad\text{and}\qquad \Su\starMP\pi_{\pm E,\pm
s}^{(MP)}(\fett{p},\fett{u}) = \pm\frac{\hbar}{2}\pi_{\pm E,\pm
s}^{(MP)}(\fett{p},\fett{u}) \text{.} \label{HSstargenvalue}
\end{equation}
The Dirac Wigner functions are idempotent: $\pi_{\pm E,\pm
s}^{(MP)}(\fett{p},\fett{u}) \starMP \pi_{\pm E,\pm
s}^{(MP)}(\fett{p},\fett{u}) = \pi_{\pm E,\pm
s}^{(MP)}(\fett{p},\fett{u})$ and with the trace (\ref{TrDef2})
the Dirac Wigner functions (\ref{pidi}) are normalized to $1$.

With the relations
\begin{equation}
[H_D,x_i]_{\starMP}=-i\hbar c\alpha_i
\qquad\text{and}\qquad\{H_D,\alpha_i\}_{\starMP}=2cp_i
\end{equation}
one can calculate
the time development of the position as
\begin{eqnarray}
x_i(t)&=&\mathrm{Exp}_{MP}(-H_D t)\starMP
x_i\starMP\mathrm{Exp}_{MP}(H_D t)\nonumber\\
&=&x_i+c^2p_it\starMP H_D^{-1\starMP}\nonumber\\
&&\qquad + \frac{i\hbar c}{2}\left(\alpha_i-cp_i \starMP
H_D^{-1\starMP}\right) \starMP
H^{-1\starMP}\starMP\left(\mathrm{Exp}_{MP}(2H_Dt)-1\right)
\text{,}\label{xi}
\end{eqnarray}
where
\begin{equation}
H_D^{-1\starMP}=\frac{H_D}{c^2\mbox{\boldmath$p$}^2+m^2c^4}
\end{equation}
is the inverse under the Moyal-Pauli star product. In Eq.\
(\ref{xi}) the first two terms correspond to the classical motion
while the last term is the well-known term that represents the
Zitterbewegung \cite{Schwabel}.

It is also possible to derive the Dirac equation in the star
product formalism by using the fact that in the rest frame it
should coincide with the $*$-genvalue Eq.\ (\ref{ham}). By setting
$\fett{p}=0$ this equation becomes
\begin{equation}
\left(\gamma^0 mc \mp mc\right)\starP\pi_{\pm E}^{(MP)}(0) = 0
\qquad\text{with}\qquad \pi_{\pm E}^{(MP)}(0)= \frac{1}{2}
\left(1\pm \gamma^0\right) \text{.} \label{rest}
\end{equation}
The solution $\pi_{\pm E}^{(MP)}(0)$ follows from
Eq.~(\ref{diracpmEpi}). According to (\ref{boost2}) the equations
in (\ref{rest}) can be boosted into a moving frame by
$S=\mathrm{Exp}_P(\fett{\omega}\cdot\fett{K})$, where the
parameter $\fett{\omega}$ depends on the momentum $\fett{p}$ of
the particle in the moving frame:
\begin{equation}
S^{-1}\starP \left(\gamma^0mc\mp mc\right)\starP\pi_{\pm
E}^{(MP)}(0)\starP S = \left(S^{-1} \starP \gamma^0 \starP S \, mc
\mp mc\right)\starP S^{-1} \ \starP \pi_{\pm E}^{(MP)}(0)\starP S
= 0. \nonumber
\end{equation}
Eq.\ (\ref{boost2}) leads to
\begin{equation}
S^{-1}\starP\gamma^0 \starP S=\frac{\psl}{mc},
\end{equation}
so that with the definition
\begin{equation} \pi_{\pm m}^{(MP)}(p)=S^{-1} \starP
\pi_{\pm E}^{(MP)}(0) \starP S
\end{equation}
the equation above turns into
\begin{equation}
\left(\psl \mp mc\right) \starMP \pi_{\pm m}^{(MP)}(p)=
0\qquad\text{with}\qquad \pi_{\pm m}^{(MP)}(p)= \frac{\pm \psl +
mc}{2mc}\text{,} \label{energie-rel}
\end{equation}
which corresponds to the Dirac equation and the well-known energy
projector respectively.

The same discussion as for the Lorentz boost of the energy
$*$-genvalue Eq.\ (\ref{ham}) can be repeated for the spin
$*$-genvalue Eq.\ (\ref{Sugl}) with its solution (\ref{Susol}). By
assuming that $\Su = \frac{\hbar}{2}\gamma_5\starP
(\fett{\gamma}\cdot \fett{u})$ is a valid spin observable in the
rest frame it takes on the form
\begin{equation}
S_u=S^{-1}\starP \Su \starP S = -\frac{\hbar}{2}\gamma_5\starP
\usl
\end{equation}
in the moving frame by applying a boost with
$S=\mathrm{Exp}_P(\fett{\omega}\cdot\fett{K})$. The condition
$\fett{u}^2=1$ and $\fett{u}\cdot\fett{p}=0$ have to be translated
into $u^\mu u_\mu=-1$ and $u^\mu p_\mu=0$ respectively to ensure
that $S_u \starP S_u=\left( \frac{\hbar}{2}\right)^2$ and
$[S_u,H_D]_{\starP}=0$ hold true in every frame. Finally, the
relativistic version of the spin $*$-genvalue equation and its
solution become
\begin{equation}
S_u \starP \pi_{\pm s}^{(P)}(u) = \frac{\hbar}{2} \gamma_5\starP
\usl \starP \pi_{\pm s}^{(P)}(u) = \pm \frac{\hbar}{2}\pi_{\pm
s}^{(P)}(u) \quad \text{with} \quad \pi_{\pm s}^{(P)}(u) =
\frac{1}{2} \pm \frac{1}{\hbar} S_u = \frac{1\mp\gamma_5\starP
\usl}{2} \label{spin-rel}
\end{equation}
by replacing $\Su$ with $S_u$ in both (\ref{Sugl}) and
(\ref{Susol}). One can see that the spin projectors $\pi_{\pm
s}^{(P)}$ take on the form which is known from the Dirac theory.
As in Eq.\ (\ref{pidi}) the two projectors in Eqs.\
(\ref{energie-rel}) and (\ref{spin-rel}) can be combined to
\begin{equation}
\pi_{\pm m,\pm s}^{(MP)}(p,u) = \pi_{\pm m}^{(MP)}(p)
\starMP\pi_{\pm s}^{(P)}(u)  = \pi_{\pm s}^{(P)}(u)
\starMP\pi_{\pm m}^{(MP)}(p)\label{pidi2} \text{,}
\end{equation}
which is a projector corresponding to the four-spinors $u$ and $v$
in the Dirac theory. It fulfills both $*$-genvalue equations in
(\ref{energie-rel}) and (\ref{spin-rel}), is idempotent, and is
normalized with respect to the trace (\ref{TrDef2}).

\section{The Non-Relativistic Limit of the Dirac Equation}
\setcounter{equation}{0}

\qquad In order to calculate the non-relativistic limit of the Dirac
Hamiltonian it is straightforward to translate the
Foldy-Wouthuysen transformation \cite{Foldy} into the star product
formalism. The time development of the Wigner function is given by
\cite{Zachos2}
\begin{equation}
i\hbar\frac{\partial\pi(t)}{\partial t}=
\left[H(t),\pi(t)\right]_{\starMP}\text{.} \label{Wigent}
\end{equation}
This can be translated into an equation for the unitary
transformed Wigner function
\begin{equation} \pi
'(t)=U(t)\starMP\pi(t)\starMP U(t)^{-1},
\end{equation}
which leads to
\begin{equation} i\hbar\partial_t\pi'(t)=\left[H'(t),\pi
'(t)\right]_{\starMP},
\end{equation}
with
\begin{equation}
H'(t)= U(t)\starMP(H(t)-i\hbar\partial_t)\starMP U(t)^{-1}.
\label{Htrans}
\end{equation}
The Hamiltonian can be written as
\begin{equation}
\frac{H}{mc^2} = \beta+\mathcal{E}+\mathcal{O}
\end{equation}
with
\begin{equation}
\beta + \mathcal{E}=\frac{1}{2}\left( \frac{H}{mc^2}+\beta
\starP\frac{H}{mc^2} \starP \beta \right) \qquad \text{and}\qquad
\mathcal{O}=\frac{1}{2}\left( \frac{H}{mc^2}- \beta \starP
\frac{H}{mc^2} \starP \beta \right) \nonumber \text{.}
\end{equation}
The function $\mathcal{E}$ has positive parity and $\mathcal{O}$
is a function with negative parity. It is assumed that $\cal{E}$
and $\cal{O}$ are of order $(\frac{1}{c})^2$ and $(\frac{1}{c})^1$
respectively.

Following the conventional Foldy-Wouthuysen procedure we
choose
\begin{equation}
U(t)={\rm Exp}_{MP}\left( \frac{i\beta}{2}\starMP \cal{O} \right)=
\sum_{n=0}^{\infty}\frac{1}{n!}\left(\frac{\beta}{2}\starMP\cal{O}
\right)^{n\starMP},
\end{equation}
so that (\ref{Htrans}) gives
\begin{eqnarray}
\frac{H'}{mc^2}&=&\beta\starMP\left(1+\frac{1}{2}{\cal{O}}^{2\starMP}
-\frac{1}{8}{\cal{O}}^{4\starMP}\right)+{\cal{E}}
-\frac{1}{8}\left[{\cal{O}},\left(\left[{\cal{O}},{\cal{E}}\right]_{\starMP}
+\frac{i\hbar}{mc^2}\dot{\cal{O}}\right)\right]_{\starMP}\nonumber\\
&&+\frac{1}{2}\beta\starMP\left[{\cal{O}},{\cal{E}}\right]_{\starMP}
-\frac{1}{3}{\cal{O}}^{3\starMP}+\frac{i\hbar}{2mc^2}\beta\starMP\dot{\cal{O}}
+\ldots \text{,} \label{entwi}
\end{eqnarray}
where the first row contains even functions only, whereas the
second row consists of odd functions only. This shows that
(\ref{entwi}) can be written as
\begin{equation}
\frac{H'}{mc^2}=\beta+ {\cal{E}}'+{\cal{O}}'.
\end{equation}
Repeating this transformation with
\begin{equation} U(t)={\rm Exp}_{MP}\left(
\frac{i\beta}{2}\starMP\cal{O}' \right)
\end{equation}
leads to
\begin{equation} \frac{H''}{mc^2}=\beta+{\cal{E}}',
\end{equation}
where all terms of the order $(\frac{1}{c})^5$ or higher are
neglected.

For the Dirac Hamiltonian $H=\fett{\alpha}\cdot\left(c\,
\fett{p}-e\fett{A}\right)+\beta mc^2+e\varphi$ we have
\begin{equation}
{\cal{E}}= \frac{e\varphi}{mc^2}\qquad\text{and}\qquad
{\cal{O}}=\fett{\alpha}\cdot\frac{c\fett{p} -e\fett{A}}{mc^2}.
\end{equation}
Up to terms of order $(\frac{1}{c})^4$ in $\frac{H''}{mc^2}$ the
transformed Hamiltonian $H''$ is therefore given by
\begin{eqnarray}
H''&=&mc^2\beta\starMP\left(1+\frac{1}{2}{\cal{O}}^{2\starMP}-\frac{1}{8}
{\cal{O}}^{4\starMP}\right)+mc^2{\cal{E}}-\frac{mc^2}{8}\left[{\cal{O}},
\left(\left[{\cal{O}},{\cal{E}}\right]_{\starMP}+\frac{i\hbar}{mc^2}
\dot{\cal{O}}\right)\right]_{\starMP}\nonumber\\
&=&\beta\left(mc^2+\frac{\left(\fett{p}-\frac{e}{c}
\fett{A}\right)^{2\starMP}}{2m}-\frac{\fett{p}^4}
{8m^3c^2}\right)-\frac{e\hbar}{2mc}\beta\starMP\fett{\sigma}\cdot
\fett{B}+e\varphi\nonumber\\
&&-\frac{e\hbar}{4m^2c^2}\fett{\sigma}\cdot\left(
\fett{E}\times\fett{p}\right)
-\frac{e\hbar^2}{8m^2c^2}\mathrm{div}\fett{E}\text{.}
\label{Hzweistrich}
\end{eqnarray}
In order to compare this result with the conventional operator
expression one has to apply a Weyl transformation $\Theta_W$
\cite{Agarwal}, which transforms a product of phase space
variables into the totally symmetrized product of the
corresponding operators and the $\sigma^i$, $\alpha_i$ and $\beta$
into the corresponding matrices. The Hamilton operator
corresponding to (\ref{Hzweistrich}) is then
\begin{eqnarray}
\hat{H}''&=&\beta\left(mc^2+\frac{\left(\hat{\fett{p}}-\frac{e}{c}
\hat{\fett{A}}\right)^2}{2m}-\frac{\hat{\fett{p}}^4}
{8m^3c^2}\right)-\frac{e\hbar}{2mc}\beta\hat{\fett{\sigma}}\cdot
\hat{\fett{B}}+e\hat{\varphi}\nonumber\\
&&-\frac{e\hbar}{4m^2c^2}\hat{\fett{\sigma}}\cdot\left(
\hat{\fett{E}}\times\hat{\fett{p}}\right)
-\frac{ie\hbar^2}{8m^2c^2}\hat{\fett{\sigma}}\cdot\mathrm{rot}
\hat{\fett{E}}
-\frac{e\hbar^2}{8m^2c^2}\mathrm{div}\hat{\fett{E}}\text{,}
\end{eqnarray}
which is the conventional result. We have used the relation
\begin{equation}
\Theta_W\left(\fett{E}\times\fett{p}\right) =
\frac{1}{2}\left(\hat{\fett{E}}\times\hat{\fett{p}}
-\hat{\fett{p}}\times\hat{\fett{E}}\right) =
\hat{\fett{E}}\times\hat{\fett{p}}+\frac{i\hbar}{2}
\mathrm{rot}\hat{\fett{E}}.
\end{equation}

\section{Conclusions}
\setcounter{equation}{0}

\qquad Starting from an underlying Grassmann algebra a process of
Chevalley Cliffordization leads to a Clifford algebra. The product
in this algebra is essentially a fermionic star product which
arises in the quantization of physical systems involving fermionic
degrees of freedom. This product is important for analysis of the
algebraic structure of quantum field theories. It also provides a
canonical procedure for quantizing physical systems with either
bosonic or fermionic degrees of freedom. The concept of spin in
relativistic and non-relativistic quantum mechanics can be
clarified in this framework.

Clifford algebras can be taken as the starting point for a
fruitful analysis of many mathematical structures which arise in
theoretical physics \cite{coll}, not only in quantum mechanics and
field theory, but also in classical mechanics \cite{Found}.
Starting from an underlying Grassmann algebra these structures in
classical mechanics may be seen as arising from a Cliffordization
procedure involving a fermionic star product. An additional
deformation of the theory by use of a Moyal star product for the
bosonic variables then leads to its quantum version. In a
subsequent paper \cite{HHS} we shall further elucidate this
unified approach for treating classical and quantum mechanical
dynamical systems.

\begin{appendix}
\renewcommand{\theequation}{\Alph{section}.\arabic{equation}}
\setcounter{equation}{0}
\setcounter{section}{1}
\section*{Appendix}

\qquad In this Appendix we show that the representation
(\ref{hakdar}) fulfills Axiom (\ref{iii}), i.e.\
\begin{equation}
(uv)\lc{B}w=u\lc{B}(v\lc{B}w).\label{iiib}
\end{equation}
Without restriction of
generality we choose $u=\theta_1\ldots \theta_r,\,
v=\theta_{r+1}\ldots \theta_s$ and $w=\theta_{i_1}\ldots
\theta_{i_t}$ with $t\ge s$. Using the abbreviations
$B(\theta_i,\theta_j)=B_{i,j}$ and $\partial_{\theta_i}=
\partial_i$ we find for the left hand side:
\begin{eqnarray}
(uv)\lc{B}w&=&uv\frac{1}{s!}\left(\sum_{i,j}B_{i,j}\lvec{\partial}_{i}
\vec{\partial}_{j}\right)^s w\nonumber\\
&=&\theta_i\ldots \theta_s\left(\sum_{\sigma\in S_{s,t}}
    B_{1,i_{\sigma(1)}}\cdots B_{s,i_{\sigma(s)}}
      (\lvec{\partial}_{1}\vec{\partial}_{i_{\sigma(1)}})\cdots
        (\lvec{\partial}_{s}\vec{\partial}_{i_{\sigma(s)}})\right)
    \theta_{i_1}\ldots \theta_{i_t}\nonumber\\
&=&\theta_i\cdots \theta_s\left(\sum_{\sigma\in S_{s,t}}
    B_{1,i_{\sigma(1)}}\cdots B_{s,i_{\sigma(s)}}
     \lvec{\partial}_{1}\cdots\lvec{\partial}_{s}
     \vec{\partial}_{i_{\sigma(s)}}\cdots\vec{\partial}_{i_{\sigma(1)}}
    \right) \theta_{i_1}\ldots \theta_{i_t}\nonumber\\
&=&(-1)^{s(s-1)/2} \sum_{\sigma\in S_{s,t}}
     B_{1,i_{\sigma(1)}}\cdots B_{s,i_{\sigma(s)}}
     \vec{\partial}_{i_{\sigma(s)}}\cdots\vec{\partial}_{i_{\sigma(1)}}
     \theta_{i_1}\ldots \theta_{i_t},
\label{A1}
\end{eqnarray}
where $S_{s,t}$ is the set of all permutations of $s$ elements out of $t$.

For the right hand side of Eq.\ (\ref{iiib}) we first calculate
\begin{eqnarray}
v\lc{B}w &=& \theta_{r+1}\cdots \theta_s \left(\sum_{\sigma\in
S_{s-r,t}}
    B_{r+1,i_{\sigma(r+1)}}\cdots B_{s,i_{\sigma(s)}}
  \lvec{\partial}_{r+1}\cdots\lvec{\partial}_{s}
  \vec{\partial}_{i_{\sigma(s)}}\cdots\vec{\partial}_{i_{\sigma(r+1)}}
\right)\theta_{i_1}\cdots \theta_{i_t}\nonumber\\
&=&(-1)^{(s-r)(s-r-1)/2}\sum_{\sigma\in S_{s-r,t}}
   B_{r+1,i_{\sigma(r+1)}}\cdots B_{s,i_{\sigma(s)}}
   \vec{\partial}_{i_{\sigma(s)}}\cdots\vec{\partial}_{i_{\sigma(r+1)}}
     \theta_{i_1}\ldots \theta_{i_t}.
\end{eqnarray}
This result lead to
\begin{eqnarray}
u\lc{B}(v\lc{B}w)&=&(-1)^{(s-r)(s-r-1)/2}\sum_{\sigma\in
S_{s-r,t}}
   B_{r+1,i_{\sigma(r+1)}}\cdots B_{s,i_{\sigma(s)}}\theta_1\cdots \theta_r
\nonumber\\
&&\times  \left[\sum_{\sigma'\in S_{r,t}}
      B_{1,i_{\sigma'(1)}}\cdots B_{r,i_{\sigma'(r)}}
  \lvec{\partial}_1\cdots\lvec{\partial}_{r}
  \vec{\partial}_{i_{\sigma'(r)}}\cdots\vec{\partial}_{i_{\sigma'(1)}}
    \right]
 \vec{\partial}_{i_{\sigma (s)}}\cdots\vec{\partial}_{i_{\sigma(r+1)}}
 \theta_{i_1}\cdots \theta_{i_t}\nonumber \\
&=&(-1)^{[(s-r)(s-r-1)+r(r-1)]/2}
     \sum_{\sigma\in S_{s-r,t} \atop \sigma'\in S_{r,t}}
      B_{1,i_{\sigma'(1)}}\cdots B_{r,i_{\sigma'(r)}}
      B_{r+1,i_{\sigma(r+1)}}\cdots B_{s,i_{\sigma (s)}}\nonumber\\
&&\qquad \qquad\qquad\qquad\qquad\qquad\qquad\qquad\times
   \vec{\partial}_{i_{\sigma'(r)}}\cdots\vec{\partial}_{i_{\sigma'(1)}}
   \vec{\partial}_{i_{\sigma(s)}}\cdots\vec{\partial}_{i_{\sigma(r+1)}}
   \theta_{i_1}\cdots \theta_{i_t}\nonumber\\
&=&(-1)^{s(s-1)/2}\sum_{\sigma\in S_{s-r,t} \atop \sigma'\in S_{r,t}}
      B_{1,i_{\sigma'(1)}}\cdots B_{r,i_{\sigma'(r)}}
      B_{r+1,i_{\sigma(r+1)}}\cdots B_{s,i_{\sigma (s)}}\nonumber\\
&&\qquad\qquad\qquad\qquad\qquad\qquad\qquad\qquad\times
   \vec{\partial}_{i_{\sigma(s)}}\cdots\vec{\partial}_{i_{\sigma(r+1)}}
   \vec{\partial}_{i_{\sigma'(r)}}\cdots\vec{\partial}_{i_{\sigma'(1)}}
   \theta_{i_1}\cdots \theta_{i_t}\nonumber\\
&=&(-1)^{s(s-1)/2}
   \sum_{\sigma\in S_{s,t}}B_{1,i_{\sigma(1)}}\cdots B_{s,i_{\sigma(s)}}
   \vec{\partial}_{i_{\sigma(s)}}\cdots\vec{\partial}_{i_{\sigma(1)}}
    \theta_{i_1}\cdots \theta_{i_t},
\label{A2}
\end{eqnarray}
which is the same result as before. In the last step we used the
fact that a term in the sum will be zero if
$\sigma(i)=\sigma'(j)$, because of the fermionic character of the
derivatives.

\end{appendix}

\end{document}